\documentclass[letterpaper,twocolumn,10pt]{article}
\usepackage{usenix-2020-09}

\usepackage{amsmath}
\usepackage{authblk}
\usepackage{booktabs}
\usepackage{makecell}
\usepackage{subfig}
\usepackage{soul}
\usepackage{tabularx}
\usepackage{tikz}

\usepackage{multicol}
\usepackage{multirow}
\usepackage{tabularx}
\newcolumntype{Z}{ >{\centering\arraybackslash}X }

\usepackage{tikz}
\newcommand*\circled[1]{\tikz[baseline=(char.base)]{
            \node[shape=circle,draw,inner sep=2pt] (char) {#1};}}
            
\newcolumntype{n}{>{\hsize=0.7\hsize\centering\arraybackslash}X}
\newcolumntype{w}{>{\hsize=1.6\hsize\centering\arraybackslash}X}

\usepackage{ifthen}
\usepackage[normalem]{ulem} 
\usepackage{xcolor}
\usepackage{amssymb}

\newboolean{showedits}
\setboolean{showedits}{true} 
\ifthenelse{\boolean{showedits}}
{
	\newcommand{\del}[1]{\textcolor{red}{\sout{#1}}} 
}{
	\newcommand{\del}[1]{} 
	
}

\newboolean{showcomments}

\setboolean{showcomments}{true}

\newcommand{\id}[1]{$-$Id: scgPaper.tex 32478 2010-04-29 09:11:32Z oscar $-$}

\ifthenelse{\boolean{showcomments}}
{\newcommand{\nbc}[3]{
 {\colorbox{#3}{\bfseries\sffamily\scriptsize\textcolor{white}{#1}}}
 {\textcolor{#3}{\sf\small$\blacktriangleright$\textit{#2}$\blacktriangleleft$}}}
 }
{\newcommand{\nbc}[3]{}
 \renewcommand{\del}[1]{} 
 }

\definecolor{swlcolor}{rgb}{0.4,0.6,0.2}

\definecolor{kbcolor}{rgb}{0.5,0,0.5}

\definecolor{sfcolor}{rgb}{0.2,0.0,0.5}

\definecolor{kscolor}{rgb}{0.9,0.4,0.6}

\definecolor{agcolor}{rgb}{0.2,0.4,0.2}

\definecolor{mccolor}{rgb}{0.21,0.54,0.84}

\newcommand{\wt}{{WiredTiger}}
\newcommand{\WT}{{\wt}~}
\newcommand{\WTs}{{\wt}}

\author[1,2]{Alexandra Fedorova}
\author[1]{Keith Smith}
\author[1]{Keith Bostic}
\author[1]{Alexander Gorrod}
\author[1]{Sue LoVerso}
\author[1]{Michael Cahill}
\affil[1]{MongoDB}
\affil[2]{University of British Columbia}

\begin{document}

\title{\Large \bf Writes Hurt:
  Lessons in Cache Design for Optane NVRAM}
\maketitle

\begin{abstract}
Intel\textregistered~Optane\texttrademark~DC Persistent Memory resides on the memory bus and approaches DRAM in access latency. One avenue for its adoption is to employ it in place of persistent storage; another is to use it as a cheaper and denser extension of DRAM. In pursuit of the latter goal, we present the design of a volatile Optane NVRAM cache as a component in a storage engine underlying MongoDB. The primary innovation in our design is a new cache admission policy. We discover that on Optane NVRAM, known for its limited write throughput, the presence of writes disproportionately affects the throughput of reads, much more so than on DRAM. Therefore, an admission policy that indiscriminately admits new data (and thus generates writes), severely limits the rate of data retrieval and results in exceedingly poor performance for the cache overall. We design an admission policy that balances the rate of admission with the rate of lookups using dynamically observed characteristics of the workload. Our implementation outperforms OpenCAS (an off-the-shelf Optane-based block cache) in all cases, and Intel Memory Mode in cases where the database size exceeds the available NVRAM. Our cache is decoupled from the rest of the storage engine and uses generic metrics to guide its admission policy; therefore our design can be easily adopted in other systems.
\end{abstract}

\section{Introduction}\label{sec:intro}


Intel\textregistered~Optane\texttrademark~DC Persistent Memory is one of the first widely available non-volatile memory (NVRAM) products, released two years prior to the time of this writing. At present the community is still grappling with the question of how to best use it in the storage stack. Although one way of adoption exploits its persistence (e.g., using it in place of another block storage device or turning applications' volatile memory into persistent), another avenue is to use it as a volatile extension to DRAM, a denser and cheaper one at that. Our study explores the second option. 

We design and implement \emph{NVCache}: an Optane NVRAM-resident volatile cache for \WT\cite{wt} -- the storage engine underlying MongoDB~\cite{mongodb}. At the heart of any cache is an admission policy. An admission policy decides, upon a cache miss, whether the missing block should be \emph{admitted}, i.e., kept in the cache after being retrieved from a lower level of storage. With few exceptions, caches indiscriminately admit data on read misses, differing only in whether they admit it on write misses. We found that such a simplistic policy decreases the throughput of write-heavy workloads up to ~80\% and read-only workloads by about ~20\%. Admitting new data into a cache generates writes -- as every newly inserted cache block  must be written into the cache memory -- and limited write throughput is a well known property of Optane NVRAM~\cite{optane-perf}. What was \emph{not} previously known was that writes to Optane NVRAM disproportionately affect the throughput of concurrent reads. While writes affect concurrent reads on any storage device, our measurements show that this effect is much larger on Optane NVRAM than on its counterpart DRAM  (see \S\ref{sec:background}). An overly eager admission rate will thus limit the rate at which existing data can be retrieved, diminishing the utility of the cache. \emph{\textbf{Admission policy must, therefore, balance between the rate of admitting new data and the rate of accessing existing data}}. Our main contribution is a new admission policy that embodies this principle.

Although our work is a case study exploring a specific point in a vast design space, our findings apply broadly to similar systems. NVCache is decoupled from the rest of the storage engine and our new admission policy relies only on the rates of data admission, removal and lookup for its decisions, so our design is easy to adopt in other storage engines or stand-alone caches. While our work addresses the idiosyncrasy of one specific storage technology, we hypothesize that the admission policy we propose will be relevant for any caching device where writes disproportionately impact reads. 

The rest of the paper is organized as follows: \S\ref{sec:background} demonstrates that writes disproportionately affect the throughput of reads on Optane NVRAM. That section also puts our work in the broader context of multi-tier caching systems, and provides relevant background on \WTs. \S\ref{sec:design} presents the basics of NVCache design, which relies on well-known methods, and then unveils the design of the new admission policy, backing its features with experimental data. \S\ref{sec:eval} compares NVCache with off-the-shelf alternatives: Intel Memory Mode~\cite{mm} and OpenCAS~\cite{opencas}, and reports the effect on performance-per-\$ of replacing part of system DRAM with Optane NVRAM. \S\ref{sec:related} describes related work and \S\ref{sec:conclusion} summarizes our findings.

\section{Background and Motivation}\label{sec:background}

\subsection{Optane memory's Achilles' heel}\label{sec:writes-suck}

Optane NVRAM has a superpower: read and write latency for small operations compete with DRAM, reads being only about 2$\times$ slower and writes being roughly the same latency as DRAM\footnote{Writes into NVRAM need only to reach the processor’s ADR (Asynchronous DRAM refresh domain).} (see \cite{optane-perf}, Fig.2). Read throughput is impressive: sequential reads reach 6GB/s per NVDIMM (see~\cite{optane-perf}, Fig.4(a)), and with a single CPU supporting up to six NVDIMMs, the throughput can climb into double digits. 

\begin{figure}[t]
\centering
\includegraphics[width=\linewidth]{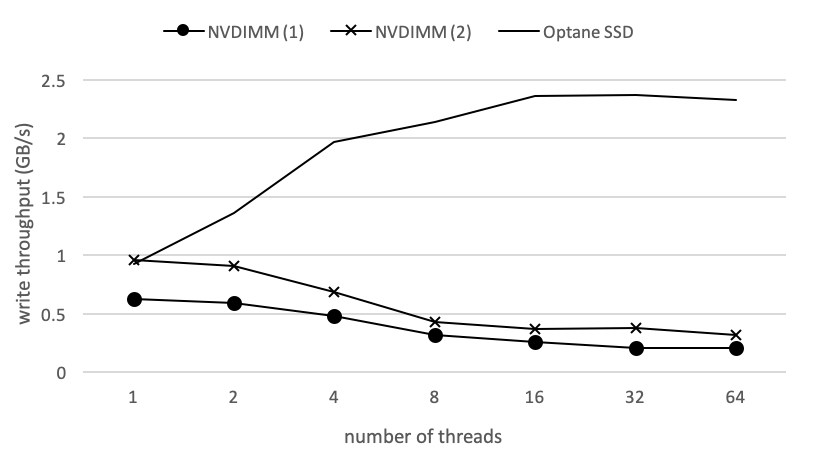}
\caption{Sequential write throughput to Optane persistent memory using one or two NVDIMMs, and to Optane SSD. Parameters of the experimental system are described in \S\ref{sec:system}.}
\label{fig:optane-writes}
\end{figure}

Optane also has an Achilles' heel: write throughput is sluggish and gets worse with many threads. Figure~\ref{fig:optane-writes} shows sequential write throughput to Optane NVDIMMs (with one and two DIMMs) and to an Optane SSD P4800X (built with the same memory technology but packaged as an SSD). Writes to Optane memory are barely competitive with the SSD using one thread, but show negative scaling as we use more threads\footnote{Our data is for non-interleaved writes. Interleaved writes will achieve higher throughput (and also negative scaling with more threads, see~\cite{optane-perf}, Fig.4(c)), but interleaving can only be used on NVDIMMs in the same NUMA node, which was not the case on our system configured according to manufacturer recommendations (\cite{intel-config}, Table 17). NVRAM access was done via \texttt{memcpy} from a \texttt{mmap}ed file residing on a DAX file system in NVRAM. This was the fastest method and it produced similar results as the fastest methods discovered by others~\cite{optane-perf}.}. 

\begin{figure}[t]
\centering
\includegraphics[width=\linewidth]{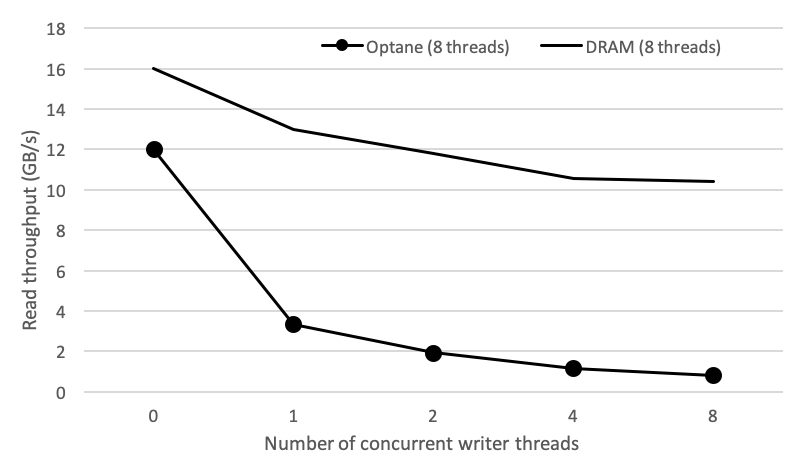}
\caption{Read throughput for Optane NVRAM (two NVDIMMs) and DRAM in with 8 reader threads and with increasing concurrent writers. The read throughput on Optane NVRAM is disproportionately affected. Parameters of the experimental system are described in \S\ref{sec:system}.} 
\label{fig:optane-reads-with-writes}
\end{figure}

This phenomenon is not new and was reported by others (see~\cite{optane-perf}, Fig.4(b)). What was \emph{not} previously shown, and is even more crucial for cache design, is that sluggish writes disproportionately affect the throughput of \emph{reads}. Figure~\ref{fig:optane-reads-with-writes} shows the read throughput on Optane NVRAM dropping precipitously in the presence of concurrent writers. Only a single concurrent writer causes read throughput to drop from a solid 12GB/s to a unimpressive 3.4 GB/s (a 72\% loss). With eight writer threads, reads proceed at only 0.8 GB/s (a 93\% slowdown)\footnote{Our system has 16 cores, so CPU contention is not the issue.}.  The same experiment on DRAM produces a milder degradation in read throughput, with a loss of only 18\% with one concurrent reader and of 35\% with eight.

The implication of this finding for cache design on Optane NVRAM is that an admission policy that eagerly accepts new data (and thus generates writes)  will disproportionately affect the speed of reads, i.e., cache lookups, severely limiting the effectiveness  of the entire system. \emph{\textbf{An admission policy, must therefore carefully balance  the rate of cache admission relative to the rate of lookups}}.

\subsection{Multi-tier caching systems}
We contribute a new design of a single-tier volatile cache in Optane NVRAM; since this cache co-exists with the DRAM cache in our storage engine (as we explain in \S\ref{sec:design}), it is helpful to discuss it in the broader context of multi-tier caches and tiered memory systems. 
Here we provide a broad overview of these areas, deferring the comparison with specific projects until \S\ref{sec:related}.

A multi-tier caching system is comprised of multiple storage devices organized as a hierarchy or a pool of caches~\cite{pavlo, heuristicdb, fb-atc21, mt-cache1, mt-cache2, mt-cache3, mt-cache4, mt-cache5, estro, dynacache, orthus, amc, amc-2, mynvm}. Tiers might include DRAM and NVRAM in front of an SSD (as in our system), a SSD in front of an HDD, or any other combination thereof, but with faster, more expensive storage generally in front of slower, less expensive storage. Studies of these systems investigate how to divide the data between the tiers to maximize performance. Broadly speaking, there are two design approaches: \emph{cooperative} and \emph{independent}. In a cooperative design the tiers are tightly coupled: one tier may evict data into another, and may inform it about the access patterns observed within its space. In an independent design each tier makes its own decisions about what data to admit and evict. There is also a middle ground, where one tier may take hints about data access characteristics from other tiers, but does not directly accept data or directives about what to cache.  Independent caches are easier to design and maintain from software engineering perspective, because they are less coupled with the rest of the system, and for this reason they are easier to port to other systems. Our design falls into the independent category, as we explain in \S\ref{sec:design}.

Multi-tier memory systems can be thought of as a sub-category of multi-tier caches, where one tier is DRAM and another is NVRAM or some other kind of slower memory~\cite{thermostat, tiered-mem-1, tiered-mem-2, tiered-mem-3, nimble, hemem, tiered-mem-pl1, tiered-mem-pl2, panthera}. These systems are typically implemented in the kernel or in a language runtime~\cite{tiered-mem-pl1,tiered-mem-pl2, panthera} and are transparent to applications. The main challenge in building them is deciding which pages must go to the ``fast'' tier and which ones to the ``slow'' tier -- the same problem that must be addressed in cooperative caches. 

Like all caches, multi-tier systems innovate on admission and eviction policies. An admission policy tells the cache when to insert new data; an eviction policy tells it which data to evict when the space becomes scarce. Typical caches always admit data on reads and vary as to whether they admit data on writes: i.e., \emph{write-allocate} or not. Multi-tier caches may also admit data as it is evicted from another tier. While most caches tune their admission algorithms to maximize the hit rate, our algorithm takes into account the \emph{rate of admission} for reasons explained in \S\ref{sec:writes-suck}. So our main contribution is the admission policy that is based on a fundamentally new principle. We believe that our innovation in admission policies will be relevant for any cache storage medium where the presence of writes disproportionately affects the throughput of reads.

\subsection{\WT}\label{sec:wt}
\WT is a persistent transactional key-value store~\cite{wt}. 
Internally it uses a B+-tree to organize the data. \WT materializes data in memory (in its DRAM cache) in a different format than it is stored on disk. Data on disk contains efficiently encoded keys and values. The keys in each block are sorted, but not indexed. When \WT reads a block from disk it decodes and indexes it, so that the data can be searched and updated efficiently. Furthermore, on-disk data may be optionally compressed and/or encrypted, and \WT decompresses and decrypts it before placing it in DRAM.

The main advantages of this two-pronged approach to data representation is that it provides efficient space utilization for stored data and fast operations for cached data. It is also the reason we adopted the independent design for our NVRAM cache, as we explain in \S\ref{sec:basics}.

\section{NVCache: a step-by-step design}\label{sec:design}

We first describe the baseline architecture of  \emph{NVCache}, which builds upon well-known techniques. Then we describe the evolution of the new admission policy design, beginning with a na{\"i}ve architecture and presenting experiments that motivate the next feature. 

\subsection{NVCache basics}\label{sec:basics}

As explained in \S\ref{sec:wt} \WT uses different formats for data stored persistently on disk and for data materialized in memory. On-disk data is stored in \textit{blocks}. In-memory data, which lives inside the engine's fixed-sized DRAM cache, is stored in \textit{pages}. Blocks contain efficiently encoded keys and values. Pages additionally contain indexing and other structures to facilitate fast operations. 

NVCache sits underneath the DRAM cache. Naturally we had to make a decision whether to use NVCache for caching pages, blocks or both. \WT already has a DRAM cache for pages, so caching pages would amount to extending the existing cache to use both DRAM and NVRAM -- a tiered cache similar to the recent one in Facebook's RocksDB~\cite{fb-atc21}. Caching blocks would entail creating a stand-alone block cache that sits between the DRAM cache and the block device. We decided to cache blocks, and not pages, for the following reasons. 

\WTs's pages are organized in memory as a B+-tree for efficient searching and updating, and pages contain pointers to other pages. If a page were to be manually copied (at application level) from DRAM to NVRAM in a tiered cache, the virtual addresses would change and any pointers would have to be updated accordingly. Updating them is an error-prone process that would require locking or other form of synchronization. \WT is lock-free on the read-path and mostly lock-free on the write path: adding synchronization would substantially compromise a core advantage of its original design. 

An alternative to implementing a tiered cache manually would be to use transparent tiered memory implemented in the kernel, such as Nimble~\cite{nimble} or HeMem~\cite{hemem}, or to build on top of CacheLib: Facebook's library for building caches that provides support for tiered memory~\cite{cachelib}. Kernel-based systems would require adopting an experimental kernel, which was not an option in a production deployment. CacheLib source became open on September 2, 2021~\cite{cache-lib-web}; building upon it is one alternative we may consider in the future, but according to the authors, CacheLib is not the best option for building a database's internal page cache, and so it could not be used as the substrate for RocksDB's page cache (see~\cite{cachelib}, Section 6 and discussion in \S\ref{sec:related}). 
Thus, for our current design we decided to use a stand-alone block cache, as it avoids the aforementioned problems, is simple to integrate in the existing storage engine and can be easily ported to other key-value stores. 

\begin{figure*}[t]
\centering
\includegraphics[width=0.9\linewidth]{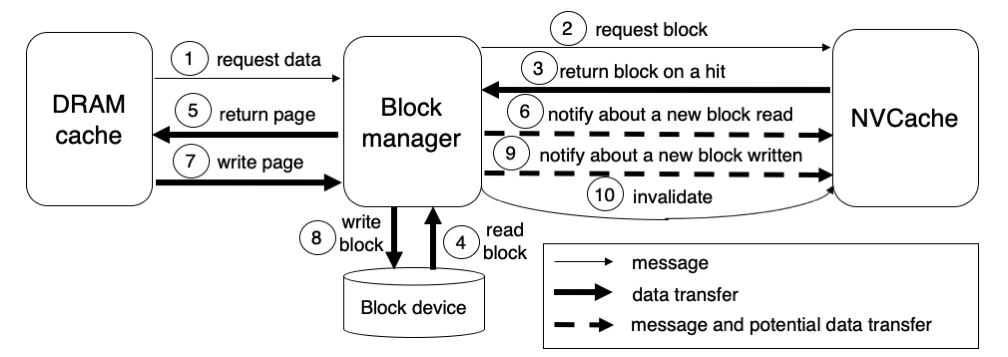}
\caption{Interaction of NVCache with the rest of the storage engine.}
\label{fig:nvcache}
\end{figure*}

 NVCache sits next to the \textit{block manager} -- the code responsible for reading/writing the data from/to disk (see Fig.~\ref{fig:nvcache}). \textbf{Read path:} If the DRAM cache cannot locate searched-for data, it issues a read to the block manager \circled{1}. The block manager checks if the block is present in NVCache \circled{2}, accessing it from NVCache if it is \circled{3} and reading it from disk if it is not \circled{4}. It then transforms the block into a page, decrypting and decompressing it if needed, and hands it over to the DRAM cache \circled{5}. If the block is not present in NVCache, NVCache has the discretion to admit it after the block manager has read it from disk \circled{6}. NVCache stores the blocks in the same format as they are stored on disk: compressed/encrypted if those configuration options were chosen. This is a feature, as storing compressed blocks increases NVRAM effective capacity.

\textbf{Write path:} The write path is not symmetrical to the read path, because \WT does not modify disk blocks in place. Updates are written into in-memory specific data structures, and then formatted into blocks and written back to disk during a process called \textit{reconciliation}. Reconciliation may occur when the DRAM cache evicts pages or as part of a database checkpoint. Reconciliation always writes a new page \circled{7}, which the block manager turns into a new block. When the block manager writes a new block \circled{8}, it notifies NVCache \circled{9}; NVCache has the discretion to admit it. Obsolete blocks are eventually freed, at which time the block manager instructs NVCache to invalidate cached copies of the freed blocks \circled{10}.

Within a broader context of multi-tier caching systems, NVCache adopts an independent design (see \S\ref{sec:background}). This is a natural consequence of our decision to cache blocks, as opposed to pages. The kernel buffer cache also caches blocks, so there is an opportunity for a cooperative design integrated with the kernel: we did not pursue this avenue, because adopting a custom kernel would not be practical in customer deployments. There are off-the-shelf NVRAM caching solutions implemented in the kernel: \textit{device mapper write cache}~\cite{dm-wc} and OpenCAS~\cite{opencas}. We describe them, evaluate OpenCAS (the more advanced of the two) and present the results in \S\ref{sec:eval}.

We experimented in the middle ground between an independent and a co-operative design, where the DRAM cache informs the NVCache on evicting a clean page (so the NVCache could bump its priority) or informs NVCache about the reason for writing a dirty block (e.g., because of  eviction or a checkpoint). Using this information did not improve NVCache effectiveness, and keeping track of it introduced overhead, so we retained a purely independent design. As a result, NVCache communicates with the block manager via a narrow API, allowing its codebase to evolve independently of the rest of the system.

Internally, NVCache is organized as a hash table with a fixed number of buckets. Upon collision, blocks mapping to the same bucket are chained in a linked list. A bucket is protected with a spinlock, but our measurements showed that the rate of collisions and the synchronization overhead were negligible (with 32K buckets for a 180GB NVCache). We use PMDK's~\cite{pmdk} allocator (based on jemalloc) to allocate NVRAM on admitting new blocks. NVCache metadata is in DRAM, but PMDK's jemalloc metadata is in the NVRAM. NVCache does not use NVRAM's persistent nature: upon exit it loses cached data. This decision simplified our design substantially, as we do not need to deal with crash consistency. The downside is that we pay the cost of re-warming the cache upon restart, and so we may revise our design in the future.  

When NVCache runs out of space it cannot admit new blocks. \emph{Eviction} is needed to purge blocks less likely to be used in order to make space for new ones. We use a simple LFRU eviction policy~\cite{lfru}. During eviction it targets blocks that were not reused within a fixed time window and evicts the least frequently used among those. Tracking of the LRU and LFU blocks is approximated so that there is no need to maintain lists. There is an eviction thread that wakes up once a second and scans the cache for eviction candidates.

\subsection{NVCache Admission Policy Design}

The NVCache admission policy is rooted in experimental data;  we therefore present the details of our experimental system and the workloads prior to exploring its design.  

\subsubsection{Experimental system}\label{sec:system}

\textbf{System:} Our system is a Lenovo ThinkSystem SR360 built with two Intel Xeon Gold 5218 processors, each having 16 hyper-threaded cores.

\textbf{Memory:} There are two Optane NVRAM modules, 126GB each, for a total capacity of 252GB. The modules are placed in separate sockets as per manufacturer recommendation. There is 196GB of DRAM; we modulate the amount available for experiments either via software (by creating a large file in ramfs) or hardware (by physically removing DRAM) in cases where the experiments demand this. We used workloads with a variety of database sizes to study conditions when the working set fits into NVRAM and when it exceeds its capacity. With a total 252GB physical capacity we are able to configure NVCache to hold at most 180GB of data. The metadata overhead of NVCache structures is kept in DRAM (and in any case it is small -- a couple of gigabytes), but the PMDK metadata, kept in the NVRAM, takes a substantial amount of space; 180GB NVCache size was the largest that we could use without experiencing out-of-memory errors from the PMDK's allocator.

\textbf{Disk:} We use Intel Optane P4800X SSD, built with the same physical media as NVRAM DIMMs, but packaged as an SSD on the PCIe bus. This SSD provides up to 2.5GB/s sequential read bandwidth and up to 2.2GB/s sequential write bandwidth. 

\subsubsection{Workloads}

While for the final evaluation (\S\ref{sec:eval}) we used the widely adopted YCSB~\cite{YCSB, ycsb-git}, during the \emph{design} process we used our in-house benchmarks. The in-house benchmarks are configuration files for a \WTs-provided workload generator application, specifying parameters such as the number of records in the database, the sizes and distributions of keys and values, the mix of operations (read, update, insert, modify, scan), the number of threads, whether or not logging and transactions are enabled, the size of the DRAM cache, the total running time, etc. The benchmarks are designed to either emulate customer workloads or to stress a particular feature (e.g., checkpoints, eviction). When presenting the throughput for a benchmark we break it down by operation type: for example, if the benchmark \textit{bm} performs a mix of reads and updates, we would report the throughput as \textit{bm.read} and \textit{bm.update}.

The workloads fall into two categories: \textbf{(1)} those that do not stand to benefit from NVCache (e.g., they use small data sets fitting entirely in DRAM, and/or they perform mostly writes) and \textbf{(2)} those that do (large data sets, read-dominant). We initially focus on benchmarks in the first category, in particular those with small data sets. The database pages are cached in the engine's DRAM cache, and its blocks -- in the kernel buffer cache as they are read from disk\footnote{\S\ref{sec:basics} explains the difference between blocks and pages.}. So even if the DRAM cache is configured to be much smaller than the dataset size, the OS buffer cache would comfortably fit blocks of small workloads. Since NVRAM caching cannot benefit these workloads, they make for an easy demonstration of the implementation overhead and are excellent workloads for exploring how to minimize it. 

\begin{figure*}[t]
\centering
\includegraphics[width=\linewidth]{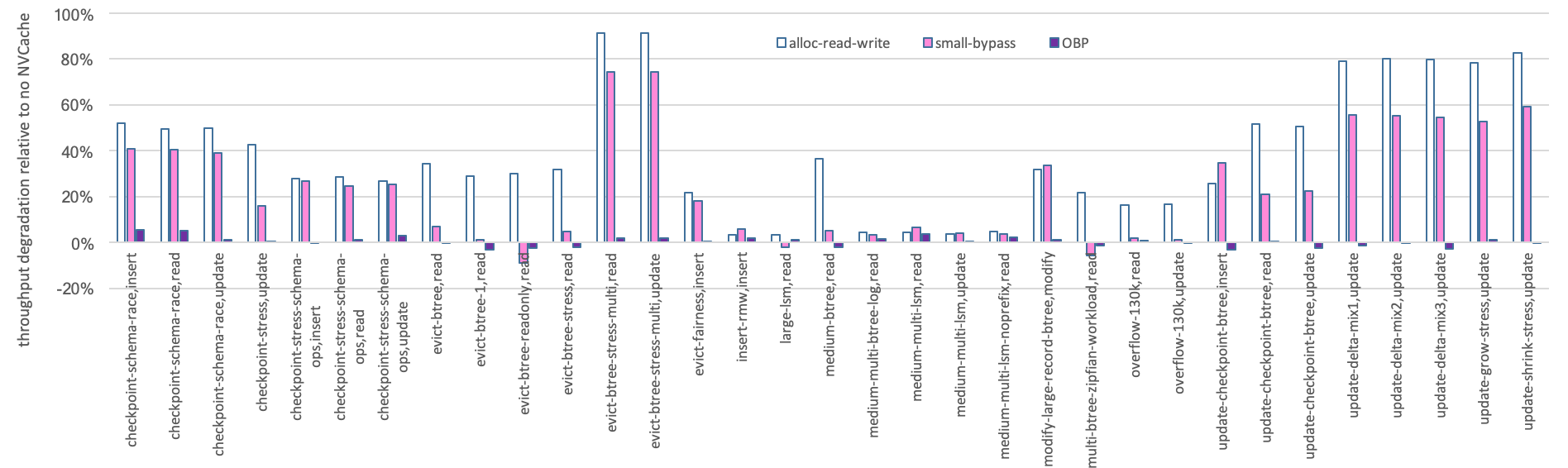}
\caption{Throughput degradation for workloads that do not stand to benefit from NVRAM caching. Lower numbers are better. Eviction is disabled during these experiments to simplify the analysis of the overhead.}
\label{fig:overhead}
\end{figure*}

\subsubsection{Lessons learned}\label{sec:lessons}
Our design rests on the three lessons that we learned in the process: (1) Bypass NVRAM for small workloads, (2) Throttle the admission rate, and (3) NVRAM cache benefit is limited to read-dominant workloads. Lesson \#2 embodies our main contribution; the others, while less novel, were also crucial for building a well-performing cache.

\paragraph{Lesson \#1: Bypass NVRAM for small datasets}

Our first and the most simple admission policy, \textit{alloc-read-write}, was always admitting a block to the NVCache when it is read from or written to disk by the block manager.  Figure~\ref{fig:overhead} shows the performance \textit{degradation} of running with 16GB DRAM and 180GB NVCache\footnote{We ran with larger DRAM sizes too, but reached the same conclusions.} for small-sized benchmarks fitting into DRAM that will not benefit from any additional caching. (Eviction is disabled during these experiments to tease apart the sources of overhead, but we re-introduce it at the end of this section.) We observe that performance penalty under this policy is substantial across the board, reaching 91\% for \textit{evict-btree-stress-multi}.

The key observation we made when analysing the causes of the overhead is that it is useless to cache data for small benchmarks that comfortably fit into DRAM -- the engine's cache or the OS buffer cache. So our first lesson is to \textbf{\textit{bypass NVCache for datasets fitting into DRAM}}. We call this feature \textit{small-bypass}, and implement it by having the NVCache monitor the aggregate size of all database files used by the workload and abstain from admitting any blocks until the dataset size outgrows the available DRAM. The bar labelled \textit{small-bypass} in Fig.~\ref{fig:overhead} shows the overhead being significantly reduced by this feature.

\textit{Small-bypass}, in a way, approximates cooperation with the OS buffer cache. NVCache cannot know which blocks the buffer cache holds, but it roughly approximates this information by juxtaposing the workload's data size and the amount of DRAM.

\begin{figure}[t]
\centering
\includegraphics[width=\linewidth]{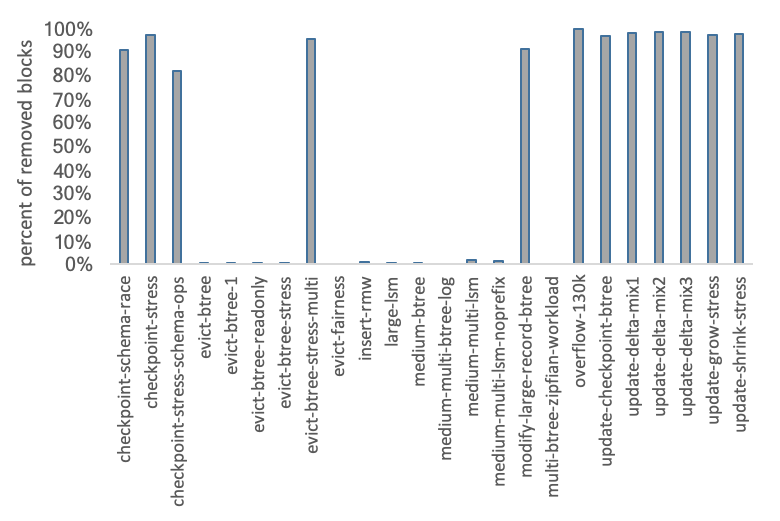}
\caption{Cached blocks that were outdated and freed. Data corresponds to the experiment in Fig.~\ref{fig:overhead}. These are aggregate data for the entire workload, so we do not show the breakdown by operation type.}
\label{fig:removed-blocks}
\end{figure}

\paragraph{Lesson \#2: Throttle the admission rate}

The \textit{small-bypass} feature all but eliminated the overhead for some workloads, but made only a small improvement for others. To show why, Figure~\ref{fig:removed-blocks} presents the number of blocks removed from NVCache because they were outdated and freed by the block manager as a percent of all admitted blocks. We observe that the benchmarks whose overhead is still substantial after the introduction of \textit{small-bypass} are those overwrite many existing blocks. 

When an application generates new data, either by inserting new key-value pairs or updating the old ones, the block manager generates new data blocks. The blocks containing old invalid data are eventually freed by the block manager and are removed from NVCache. Removing a block from NVCache involves freeing its associated memory in NVRAM, and since the PMDK allocator keeps its metadata in NVRAM, freeing a block generates writes into NVRAM. Moreover, removing old blocks creates space for new blocks, and NVCache eagerly admits data in the freed space. That also generates writes. As we showed in \S\ref{sec:background} writes disproportionately affect the throughput of reads, i.e., of cache lookups.

One could simply disable the cache for write-intensive workloads, but even read-dominant workloads will suffer from the interference of writes if overly eager eviction makes it possible to admit blocks at a high rate. Admitting new blocks generates writes, and writes will interfere with reads. 

Consider data in Table~\ref{tab:aggressive-eviction} for the three read-dominant workloads from Table~\ref{tab:large-workloads} (this table contains workloads with large working sets, for which caching may be beneficial). Table~\ref{tab:aggressive-eviction} shows data for experiments with eviction configured to eagerly evict unused blocks and for experiments configured to run without any eviction at all. When the cache is full and new blocks cannot be admitted, eager eviction frees up the space. While admitting recently referenced blocks in favour of those that were less recently accessed should improve the hit rate, this also generates more writes into NVRAM, which may diminish the rate at which we can read cached blocks. Indeed, we see from Table~\ref{tab:aggressive-eviction} that even though the cache hit rate with eager eviction is higher (as expected), the overall throughput is substantially lower than without any eviction. That is because the amount of cache writes produced with eviction is substantially higher than without it, and the writes slow down the reads. 

\begin{table}[h]
    \centering
    \begin{tabularx}{\columnwidth}{|Z|Z|Z|Z|Z|}
    \toprule
        \textbf{} &  \multicolumn{2}{|c|}{\textbf{Eager eviction}}  & \multicolumn{2}{|c|}{\textbf{No eviction}}  \\ \midrule
        \textbf{WL} & \textbf{ops/sec} & \textbf{hit rate} & \textbf{ops/sec} & \textbf{hit rate} \\
        \midrule
        \emph{evict-btree-large} & 61,699 & \textbf{48\%} & \textbf{162,690} & 44\% \\
        \hline
        \emph{evict-btree-scan.read} & 97,491 & \textbf{45\%}  & \textbf{134,404} & 36\% \\
        \hline
        \emph{medium-btree-large} & 62,012 & \textbf{48\%} &  \textbf{164,644} & 44\% \\
        \bottomrule
    \end{tabularx}
    \caption{Throughput of read-dominant workloads suffers substantially with aggressive eviction despite it producing a higher cache hit rate. Aggressive eviction generates many writes that hurt the throughput of cache lookups (reads).}
    \label{tab:aggressive-eviction}
\end{table}

The question we then ask is: \emph{\textbf{how to balance the rate of block admission and removal, which generate writes, with the rate of cache lookups, which produce reads?}} To address it, we introduce the \textit{overhead bypass} ratio (OBP):

\begin{center} 
$OBP=\frac{blocks\_inserted + blocks\_removed}{blocks\_looked\_up}$
\end{center}

Intuitively, the quantity in the numerator captures the cost of using the cache: the write-generating insertions and removals. The quantity in the denominator captures the benefit: cache lookups. OBP thus expresses the balance between the cost and benefit of using the cache; we experimentally determined that a target ratio of 10\% works best, but  settings between 5\% and 30\% were also acceptable. If OBP were to be ported and tuned for different hardware, the thresholds would be adjusted according to the degree to which concurrent writes affect the reads. E.g., on hardware where writes have a smaller effect on the performance of reads, acceptable OBP thresholds would be higher. 

NVCache continuously updates OBP and abstains from admitting or evicting cache blocks if OBP exceeds its target (10\%). The OBP metric proved remarkably stable across workloads and cache sizes. We also found OBP to work better than a simple no-write-allocate policy or OBP used in conjunction with the no-write-allocate policy. The \textit{small-bypass+OBP} bar in Figure~\ref{fig:overhead} shows that \textit{small-bypass} and OBP completely eliminate the overhead for the benchmarks that do not stand to benefit from caching. 

\paragraph{Lesson \#3: Only read-dominant workloads benefit}

While the previous sections focused on the overhead and thus experimented with small-sized workloads that do not stand to benefit from NVCache, here we switch to using large-sized workloads, which teach us the third lesson: NVRAM cache benefits only read-dominant workloads. Prior study of a custom NVRAM cache for Facebook's RocksDB came to a similar conclusion~\cite{fb-atc21}.

\begin{figure}[t]
\centering
\includegraphics[width=\linewidth]{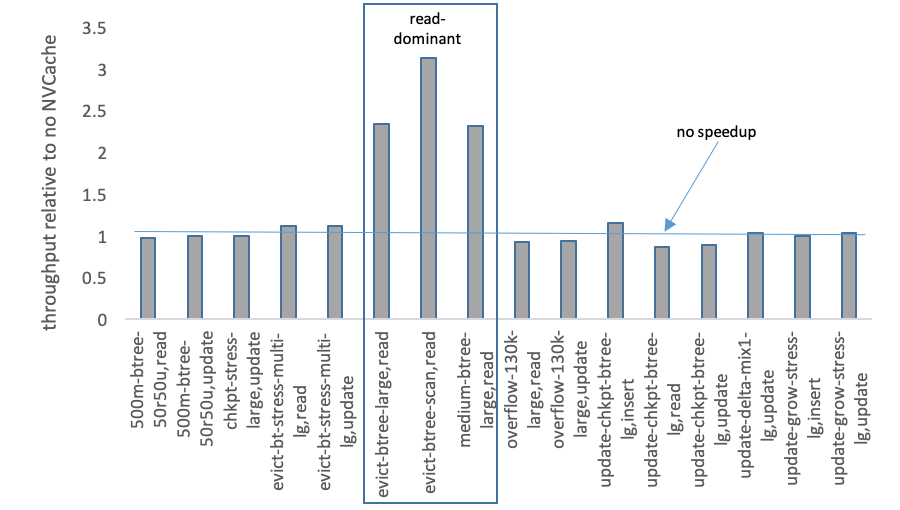}
\caption{Workloads with large datasets. 32GB DRAM and 180GB NVCache.}
\label{fig:large-workloads}
\end{figure}

%
\begin{table*}[t]
	\centering
\begin{tabularx}{\textwidth}{*{7}{|Z}|}
\toprule
		\textbf{Workload} 
		& \textbf{Op mix (threads), data size} 
		& \textbf{DRAM cache size} 
		& \textbf{NVCache hit ratio} 
		& \textbf{Removed / inserted ratio} 
		& \textbf{Amount of data written to SSD} 
		& \textbf{Amount of data admitted to cache} \\
		\midrule
\textbf{500m-btree-50r50u}  & 50\% read, 50\% update (20), 163GB & 28GB & 6\% & 98\% & 2190GB & 191GB 
\\ \hline
\textbf{chkpt-stress-lg} & 100\% update (6), 134GB & 28GB & 2\% &  94\% & 780GB & 36GB 
\\ \hline
\textbf{evict-bt-stress-multi-lg} & 80\% read, 20\% update (100), 250GB & 1GB & 20\% & 94\% & 1740GB & 424GB 
\\ \hline
\textbf{evict-btree-large} & \textbf{100\% read} (16), 120GB & 28GB & \textbf{97\%} & 0\% & 120GB & 115GB 
\\ \hline
\textbf{evict-btree-scan} & \textbf{95\% read}, 4\% insert*, 1\% update* (430), 250GB  & 28GB & \textbf{97\%} &  47\% & 400GB & 300GB 
\\ \hline
\textbf{medium-btree-large} & \textbf{100\% read} (16), 120GB & 28GB & \textbf{97\%} &  0\% & 120GB & 115GB 
\\ \hline
\textbf{overflow-130k-lg} & 50\% read, 50\% update (20), 253GB & 21GB & 6\% &  95\% & 2000GB & 127GB 
\\ \hline
\textbf{update-chkpt-btree-lg} & 90\% insert, 5\% read, 5\% update
(5), 185GB & 25GB & 6\% &  95\% &1720 GB & 137GB 
\\ \hline
\textbf{update-delta-mix1-lg} & 100\% updates (6), 125GB & 20GB & 2\% &  98\% & 2000GB & 93GB 
\\ \hline
\textbf{update-grow-stress-lg} & 96\% update, 4\% inserts* (5), 190GB & 20GB & 2\% &  97\% & 2100GB & 98GB
\\
\bottomrule
\end{tabularx}
  \caption{Properties of `large' workloads. }
  \label{tab:large-workloads}
\end{table*}

Table~\ref{tab:large-workloads} shows the large-sized workloads and their characteristics. The rate of operations marked with an asterisk (e.g., \emph{insert}, \emph{update} for \emph{evict-btree-scan}) is kept constant by the workload generator, and so we do not report their throughput, because it is largely insensitive to the system configuration. The \textit{data size} reported in the second column is the on-disk size of the database reported at the end of the run. The intermediate database size may be much larger at points when many new blocks were written to disk, but the outdated ones were not yet freed. Column six reports the total amount of data written to SSD during the run. This amount is non-zero even for read-only workloads, because it includes the data written to populate the database prior to the measured benchmark run. Although NVCache is enabled during the populate phase, it hardly admits any blocks, because OBP throttles the admission rate during this write-only phase. So when the measured run begins, NVCache is empty; it warms up during the measured run. All benchmarks run for 60 minutes, with the exception of \emph{500m-btree-50r50u}, which runs for 120.

Figure~\ref{fig:large-workloads} presents the throughput of large workloads with 32GB DRAM and 180GB of NVCache. (Data with other memory sizes leads to similar conclusions, so we omit it.) 

Read-intensive workloads benefit from NVCache substantially, running over 3$\times$ faster with the cache than without it (e.g., \textit{evict-btree-scan,read}). But even a small proportion of writes  substantially limits performance potential: \textit{evict-btree-stress-multi} performs 20\% of write operations, but the performance boost it gets from NVCache is only 12\%.

Write-intensive workloads do not benefit from NVCache, nor would they benefit from any sort of block caching, because the churn that they generate, continuously adding and removing blocks, makes most of the cache content outdated. Table~\ref{tab:large-workloads} shows the NVCache hit ratio and the fraction of removed blocks relative to those inserted. The data tells us two things: (1) workloads that don’t benefit from the cache have a very low hit ratio, (2) the low hit ratio is likely because they remove most of the blocks they insert. They write terabytes of data throughout the run (Column 6), even though their database size at the end of the run is no larger than a couple hundred gigabytes (Column 2), overwriting most of the data that they generate.

These data suggest that admitting zero blocks for write-dominant workloads would be the most practical strategy, but since the degree of write-intensity is not always known \textit{a priori}, we rely on the OBP feature to limit the damage. As Figure~\ref{fig:large-workloads} shows, OBP effectively prevents performance overhead for write-dominant workloads, and columns 6 and 7 of  Table~\ref{tab:large-workloads} show that OBP filters the majority of the write traffic to NVRAM. 

As we explained earlier, \WT does not update existing blocks in place, so a write-dominant workload will most certainly invalidate old blocks. A storage engine that does update  data in-place may be less sensitive to the phenomenon described in this section. However, given a limited write throughput of Optane NVRAM and given that the RocksDB study~\cite{fb-atc21} reached a similar conclusion, we expect the lesson learned here to be broadly applicable.

\subsubsection{Summary}
We presented three lessons in design of caches residing in Optane NVRAM: 
\begin{enumerate}
    \item Detect workloads that fit into the OS buffer cache and do not admit their blocks.
    \item Admitting blocks into Optane NVRAM produces writes, which slow down the reads, i.e., cache lookups. The admission policy must balance the cost of admitting data into the cache against the benefit of using it later.
    \item Optane NVRAM caches benefit read-dominant workloads. For write-dominant workloads, the admission policy must minimize the number of admitted blocks.
\end{enumerate}

Our admission policy uses the \textit{small-bypass} feature to embody the first lesson, and the OBP feature to embody the second and third.

\section{Evaluation}\label{sec:eval}

We evaluate NVCache using the YCSB benchmarks~\cite{ycsb-git, YCSB}. We tuned the algorithms and parameters of the NVCache using only our in-house benchmarks (a ``training set'', to use an analogy from statistical modeling), and performed no additional tuning during this final evaluation phase, using YCSB as the ``test set''. We ran experiments on the system described in \S\ref{sec:system}, varying the amount of DRAM and NVRAM. Parameters of the YCSB benchmarks are shown in Table~\ref{tab:YCSB}\footnote{We did not include YCSB-F: it is modify-heavy, and modify operations in our storage engine were designed to trade performance for smaller cache footprint and smaller log records. Therefore, the overall throughput in modify operations was very low and insensitive to memory configurations.}. The DRAM cache size was set to half of the available DRAM\footnote{The engine's cache and the OS buffer cache share the available DRAM, so this setting gives each an equal share.}, but capped at 40GB.

\begin{table}[h]
    \centering
    \begin{tabularx}{\columnwidth}{|n|w|n|}
    \toprule
        \textbf{Workload} & \textbf{Op mix, threads} & \textbf{Dataset} \\
        \midrule
        YCSB-A & 50\% read, 50\% update, 20 & 130GB \\
        \hline
        YCSB-B & 50\% read, 50\% update, 20 & 194GB \\
        \hline
        YCSB-C & 100\% read, 20 & 259GB \\
        \hline
        YCSB-D & 95\% read, 5\% insert, 100 & 219GB \\
        \hline
        YCSB-E & 95\%  scan, 5\% insert, 20 & 210GB \\
        \bottomrule
    \end{tabularx}
    \caption{YCSB characteristics}
    \label{tab:YCSB}
\end{table}

Our evaluation asks two questions:
\begin{enumerate}
    \item How does NVCache compare to off-the-shelf solutions pursuing similar goals?
    \item What is the effect of using an NVRAM cache on performance-per-\$?
\end{enumerate}

\subsection{Comparison with off-the-shelf solutions}\label{sec:mm-opencas}

\subsubsection{Baselines used for comparison}\label{sec:baselines}

We compare with two solutions that permit using NVRAM as an extensions of DRAM, available in off-the shelf Optane systems: \emph{Intel Memory Mode} (MM)~\cite{mm} and \emph{Intel Open Cache Acceleration Software} (OpenCAS)~\cite{mm}. For potential future deployment of NVRAM in the field, it was important for us that these alternatives were available in standard Linux servers and did not require custom unsupported kernels.

\textit{Intel Memory Mode} is a hardware configuration that presents Optane NVRAM to the rest of the system as regular volatile memory, and uses DRAM transparently as its cache, with data transferred between the two in units of cache lines. This is an attractive alternative, because it permits using NVRAM as an extension to DRAM without requiring any code changes, and makes it available for all data structures, in user space and kernel alike. In contrast, NVCache makes NVRAM available only for caching database file blocks. 

Memory Mode can be enabled only in specific hardware configurations (\cite{intel-config}, Table 17). We were able to successfully configure MM such that each NVDIMM was ``paired'' with a DRAM DIMM, meaning that it must be placed in the unused slot of the same channel of the same iMC (integrated memory controller) as the NVDIMM. Using additional DRAM DIMMs that were not paired with NVDIMMs produced configuration errors on our system, so we could only use the configuration with two NVDIMMs and two DRAM DIMMs. Our DRAM DIMMs were 16GB in size, so that restricted us to a configuration with 32GB of DRAM. Fortunately, MM could be configured to use all or part of the NVRAM, so we were able to vary the amount of NVRAM in the experiments.

In MM, the amount of total system memory is reported to be the same as the size of the NVRAM dedicated to MM. Since the \WTs's DRAM cache is always configured to be half the size of the \emph{physical} DRAM (see the beginning of \S\ref{sec:eval}) for equitable comparisons other systems, the kernel buffer cache will dynamically expand to use more plentiful system memory as the NVRAM size grows.  So in essence our configuration with MM uses NVRAM in the same way as NVCache does (for caching file blocks), but via an off-the-shelf hardware solution and without any code changes. An alternative would be to increase the size of the engine's DRAM cache as NVRAM grows; exploring this option in-depth was difficult due to space constraints, and thus was deemed outside the scope of the current work.

\textit{OpenCAS} is an open-source software project supported by Intel that allows using a fast block device as a cache for a slow block device, and it can be configured so that NVRAM acts as a block cache for the SSD -- same idea as NVCache. OpenCAS can be configured in several modes~\cite{opencas-config}: \emph{write-back}, \emph{write-through}, \emph{write-around}, \emph{pass-through} (disabled) and \emph{write-only} (allocate blocks only on write). Based on the lessons learned during admission policy design,  \emph{write-around} seemed the most appropriate configuration option: ``\textit{In write-around mode, the caching software writes data to the flash device if and only if that block already exists in the cache and [...]  further optimizes the cache to avoid cache pollution in cases where data is written and not often subsequently re-read.}''~\cite{opencas-config}

\textbf{Alternative baselines not pursued:} Other alternatives to compare would be device mapper write cache (\emph{dm-wc})~\cite{dm-wc} and First Responder~\cite{fr} -- both OS-level block caches, and tiered memory systems, such as Nimble~\cite{nimble} and HeMem~\cite{hemem}. We considered comparing to \emph{dm-wc} (the source code for First Responder is not available at the time of the writing), but upon analysing its properties we discovered that \emph{dm-wc} admits blocks only on writes and does not throttle the admission rate, which runs counter to the lessons learned in our design. For example, \emph{dm-wc} would admit zero blocks for read-only workloads, depriving of NVRAM caching the very workloads that benefit the most. OpenCAS, in contrast, can be configured with flexible admission policies, superseding \emph{dm-wc} in that regard. 

Nimble~\cite{nimble} and HeMem~\cite{hemem} are tiered memory systems that transparently move application pages between DRAM and NVRAM depending on how the pages are accessed. We did not compare against them, because they both required custom kernels, which would be impractical for us to adopt in the field. Furthermore, HeMem uses the NVRAM tier only for large allocations exceeding 1GB (HeMem specifically targets ``big data'' systems), so it would not use NVRAM for our engine's pages or blocks, whose size is on the order of a dozen kilobytes. 

\subsubsection{Results}

\begin{figure*}[t]
\centering
\subfloat[64GB NVRAM]{\includegraphics[width=0.33\textwidth]{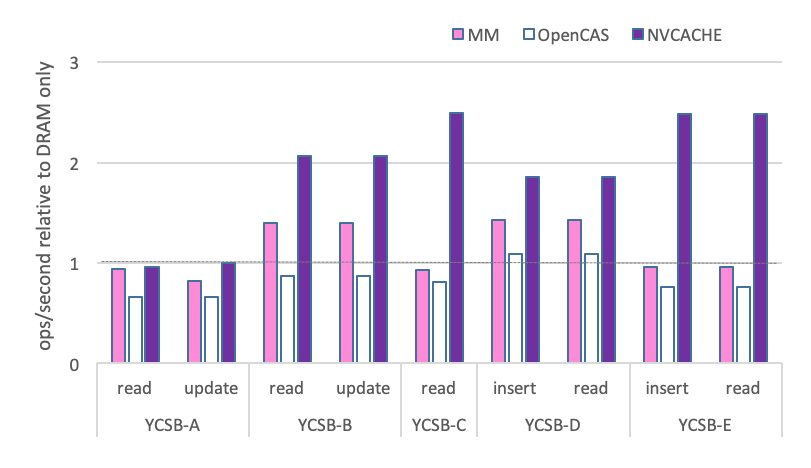}}
\subfloat[128GB NVRAM]{\includegraphics[width=0.33\textwidth]{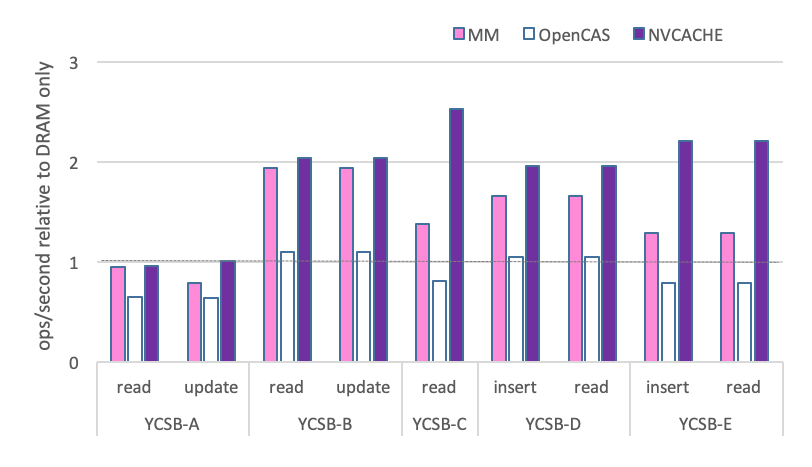}}
\subfloat[252GB NVRAM]{\includegraphics[width=0.33\textwidth]{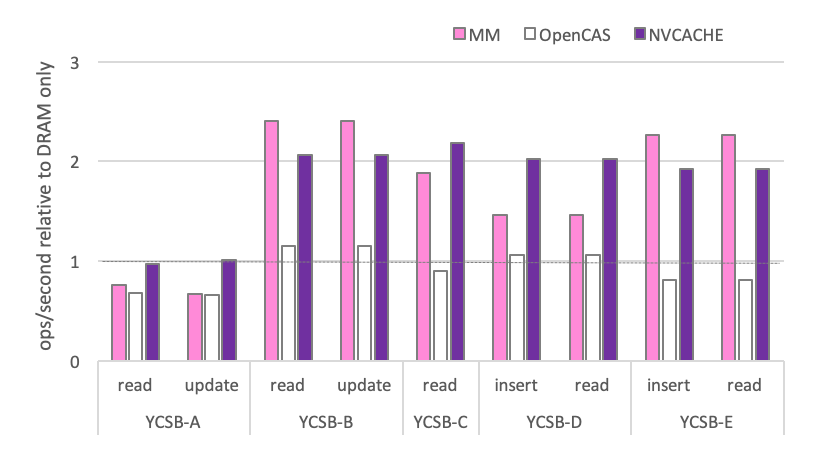}}
\caption{YCSB throughput under memory mode, OpenCAS and NVCache relative to 32GB DRAM and zero NVRAM.}
\label{fig:ycsb-mm-opencas}
\end{figure*}

Figure~\ref{fig:ycsb-mm-opencas} shows the throughput of the memory mode (MM), OpenCAS and NVCache with 32GB DRAM and 64GB, 128GB and 252GB of NVRAM relative to using no NVRAM at all. We make the following observations: 

\textbf{Observation 1:} \ul{OpenCAS cache derives no performance benefit from NVRAM.} This occurs, we conjecture, because it does not throttle the admission rate. OpenCAS delivers similar or better read hit rate as the NVCache (numbers not shown), but also makes two orders of magnitude more writes to NVRAM. With these observations, our best explanation is that \textbf{\textit{failing to throttle the admission rate to NVRAM is the main reason why OpenCAS fails to perform}}.	

\textbf{Observation 2:} \ul{Memory mode outperforms or performs comparably to NVCache when NVRAM is ample}, as shown in  Figure~\ref{fig:ycsb-mm-opencas}(c). The amount of NVRAM available for the experiments in Figure~\ref{fig:ycsb-mm-opencas}(c) is 252GB; given the dataset sizes in  Table~\ref{tab:YCSB} we observe that they, for the most part, comfortably fit into the NVRAM. System memory is reported to be 252GB in Memory Mode, and as a result the kernel buffer cache has ample capacity to expand into the NVRAM space, providing no competition for the engine's DRAM cache. By contrast, with NVCache the amount of system memory is 32GB, and the kernel buffer cache competes with the engine's DRAM cache, swapping some of its pages to disk. At the same time we observe that for NVCache the marginal utility of additional NVRAM is small after it reaches 128GB. E.g., increasing the available NVRAM from 64GB to 128GB, NVCache hit rate grows by about 20\%, but going from 128GB to 252GB, it grows by only another 5\%. 

On the other hand, we observe that Memory Mode hurts performance of the write-intensive YCSB-A (by about 30\%), while NVCache keeps it unchanged.

\textbf{Observation 3:} \ul{When the dataset size exceeds NVRAM capacity, NVCache provides substantially better performance than Memory Mode}. As shown in Fig.~\ref{fig:ycsb-mm-opencas}(a), NVCache outperforms the memory mode by between 30\% (for YCSB-B) and 169\% (YCSB-C). Further, the memory mode hurts YCSB-A’s update throughput by about 18\% relative to the DRAM-only baseline, while NVCache doesn’t. We conclude that a bespoke cache can be superior to Memory Mode when the dataset size substantially exceeds the available NVRAM.

\subsubsection{Combining Memory Mode and NVCache}
We also experimented with configurations where part of the NVRAM is dedicated to MM and the remainder is used in AppDirect mode for NVCache, reasoning that we could size NVCache such that its marginal utility is highest (~128GB), and the rest of the NVRAM could be used as MM's system memory for the benefit of other data structures. Unfortunately, we observed orders of magnitude worse throughput than with either MM or NVCache alone, and did not pursue this avenue further.

\subsection{Performance vs. cost}

\begin{figure*}[!tbp]
\centering
\subfloat[Performance per \$]{\includegraphics[width=0.55\textwidth]{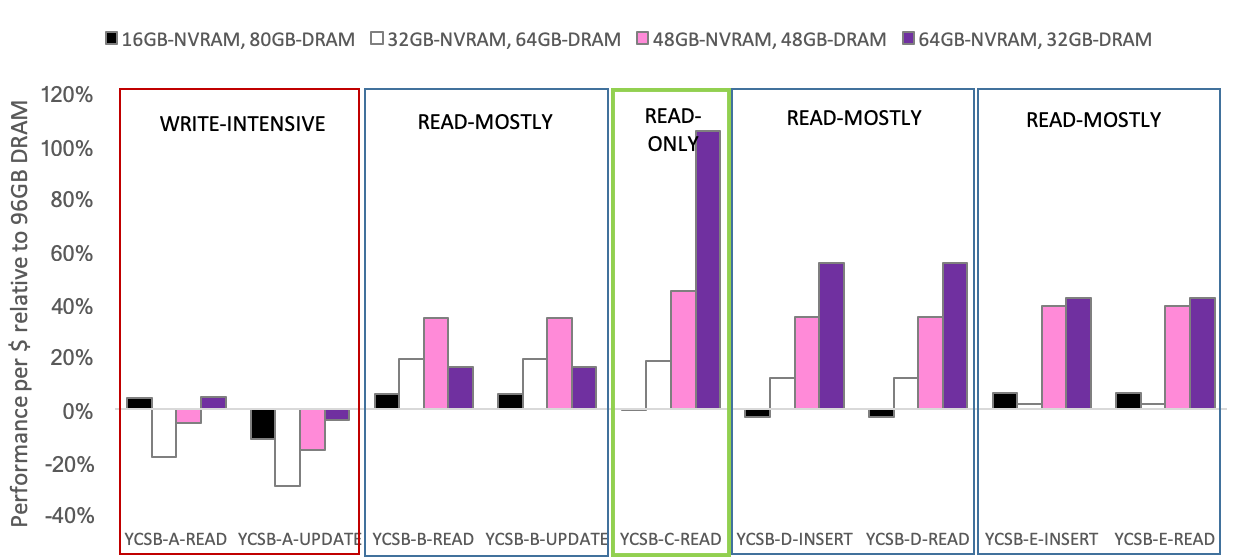}}
\subfloat[NVCache hit ratio]{\includegraphics[width=0.45\textwidth]{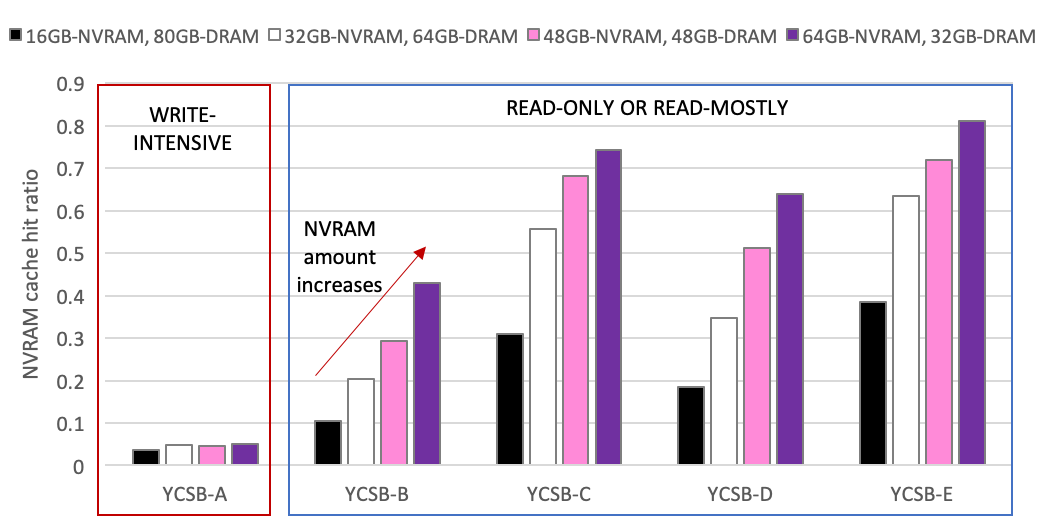}}
\caption{YCSB performance per dollar and NVCache hit ratio under increasing NVRAM and decreasing DRAM.}
\label{fig:ycsb-perf-cost}
\end{figure*}

In this experiment we take a fixed memory budget of 96GB and vary the fraction used by DRAM and NVRAM as shown in  Table~\ref{tab:perf-cost-config}\footnote{We do not use the configuration with 16GB DRAM, because a scarce DRAM amount triggered a known kernel bug in the DAX code (at fs/inode.c:530).}. We perform the experiments in this section using only NVCache, as we are unable to vary the amount of DRAM used in MM (see \S\ref{sec:baselines}) and OpenCAS proved to be not competitive.

\begin{table}[h]
    \centering
    \begin{tabularx}{\columnwidth}{*{3}{|Z}|}
    \toprule
        \textbf{NVRAM} & \textbf{DRAM} & \textbf{Relative cost} \\
        \midrule
        0GB & 96GB & 1 \\
        \hline
        16GB & 80GB & 0.90 \\
        \hline
        32GB & 64GB & 0.79 \\
        \hline
        48GB & 48GB & 0.69 \\
        \hline
        64GB & 32GB & 0.59 \\
        \bottomrule
    \end{tabularx}
    \caption{NVRAM and DRAM amounts and the cost of all system memory relative to an all-DRAM setup.}
    \label{tab:perf-cost-config}
\end{table}

We use the NVRAM/DRAM per-byte cost ratio of 0.38, same as in a recent study with Optane memory~\cite{fb-atc21}. As the amount of NVRAM increases and the amount of DRAM decreases, the total cost of system memory also decreases, as shown in Column 3. 

Figure~\ref{fig:ycsb-perf-cost}(a) shows the performance of YCSB normalized to the 96GB DRAM configuration and divided by the cost ratio in Column 3. In other words, these are performance/\$ numbers relative to the DRAM-only configuration. Positive numbers mean that the performance decreased less than the memory cost. Read-only or read-mostly workloads that benefit from the NVCache (see cache hit ratios in Fig.~\ref{fig:ycsb-perf-cost}(b)) experience a positive gain, as expected. 

While in most cases performance predictably drops as the amount of DRAM decreases, YCSB-C in configuration with 64GB NVRAM and 32GB DRAM actually performs better than it does with 96GB DRAM -- so we decrease the system cost \emph{and} improve performance in absolute terms. This occurs because going beyond 32GB DRAM the utility of additional DRAM (and a larger DRAM cache) is considerably smaller than the loss in performance due to a smaller NVCache.

YCSB-A, whose write intensity makes it unable to benefit from any additional caching, suffers the overall loss in terms of performance/\$, as its performance drops at a steeper rate than the memory cost as we decrease the amount of DRAM.

\subsection{Summary}

Our evaluation revealed that the memory mode is a competitive off-the-shelf alternative to a custom NVRAM cache when the amount of NVRAM is ample, but when it is scarce a custom cache solution such as NVCache will deliver better performance. OpenCAS is not competitive with either NVCache or the memory mode.

NVRAM is a cost-effective method of reducing memory cost while balancing the impact on performance for read-dominant workloads, where in some cases we can both reduce cost \emph{and} improve performance as DRAM is swapped in favour of NVRAM. For write-intensive workloads, however, replacing part of DRAM with NVRAM is not a cost-effective option. 
\section{Related Work}\label{sec:related}

The most similar and recent counterpart to our study is a volatile Optane-resident cache for Facebook's RocksDB~\cite{fb-atc21}. That work takes RocksDB's DRAM block cache and turns it into a two-tiered cache of DRAM and NVRAM, making it similar to tiered memory systems. Like other tiered memory systems, it addresses the question of how to split the cached data between the DRAM and the NVRAM tiers. We present a different design, that uses a stand-alone block cache interposed between the DRAM cache and the block devices.  Although the RocksDB study also uses Optane NVRAM as the cache media, it does not raise awareness about the detrimental impact of concurrent writes on reads -- a new finding we share -- and does not factor it into the admission policy.

 HeuristicDB~\cite{heuristicdb} is a cooperative block layer cache that uses a fast Optane SSD as a caching tier in front of a slower drive. HeuristicDB admits all blocks read from and written to the block device, except those part of sequential access pattern. While this generous admission policy might work for Optane SSDs, we demonstrated that it is unacceptably costly for Optane NVRAM.
 
 MyNVM is another key-value store based on RocksDB that uses Optane SSD as the medium for an internal block cache~\cite{mynvm}. Similarly to NVCache, MyNVM caters its admission policy to the properties of the Optane device, but pursues different goals: (1) to extend its endurance MyNVM admits only carefully selected keys, and (2) to maximize its bandwidth it accumulates keys into relatively large 128KB blocks before writing them to the device. While we did not focus on improving endurance, write throttling performed in NVCache via the overhead bypass parameter (OBP -- see \S\ref{sec:lessons}) should increase it. The second goal is potentially applicable to our NVRAM device too, but prior experiments (\cite{optane-perf}, Fig. 5) showed that writing into Optane NVRAM blocks larger than those that we already write (e.g., 16KB+) does not improve its bandwidth.
 
Like MyNVM, Facebook's CacheLib caters to the properties of (flash) SSDs by limiting the admission rate to promote device endurance~\cite{cachelib}. The throttling heuristic is rather simple; it is a configurable probability $p$ that determines the overall rate of admission. As far as we know, $p$ is not dynamically adjusted based on the observed rates of writes and lookups, so it could not be used in place of NVCache's OBP.

The work by Arulraj et al.~\cite{pavlo} establishes a broad framework for reasoning about multi-tiered caching systems comprised of DRAM, persistent memory and SSDs. The authors propose an algorithm for data placement that dynamically tunes the following probabilities: the probability of bypassing DRAM on reads and writes (data being read/written directly from/to NVRAM) and the probability of bypassing the NVRAM on reads and writes. Bypassing DRAM is not applicable in our engine, because DRAM stores data in a different format than NVRAM, but bypassing NVRAM is the same question we grappled with during the design of our admission policy. Arulraj's work uses simulated annealing to dynamically adapt these probabilities, while we use a dynamically computed OBP threshold. Their evaluation was performed on a simulator, while we used real hardware, which revealed concrete limitations and influenced our design. 

Estro et al. explored the relationship of performance and cost and the effects of different cache settings (such as write-through vs. writeback) in multi-tier caching configurations on real hardware~\cite{estro}. Performing similar analysis would be a natural extension of our work, but can only be done after understanding the idiosyncrasies of cache design using recently adopted memory technology, contributed by our study. 

The design of Orthus~\cite{orthus} was driven by an observation similar to ours: a seemingly faster device (Optane SSD, in their case) outperforms a slower device (a flash-based SSD) in general, but lags behind it under high concurrency. Orthus embraces a hybrid design: initially, a faster device acts as a cache for a slower device, admitting all blocks until a desired hit rate is accomplished. Then Orthus switches to a ‘’tiered mode’’, where the load is distributed among both devices to maximize the overall throughput. Our OBP feature accomplishes a somewhat similar effect when it begins throttling the admission rate to NVCache, and as a result more reads are being sent to the storage device over time. In contrast to Orthus, NVCache throttles the admission rate based on the observed cost/benefit metric, and not as a consequence of achieving a certain hit rate. In fact, we observed that it may be beneficial for overall performance to throttle the admission rate at the expense of the reduced hit rate.

Multi-tiered memory systems dealt primarily with the policies for selecting the right tier for a memory page, and (to that end) efficiently tracking page access patterns ~\cite{thermostat, tiered-mem-1, tiered-mem-2, tiered-mem-3, nimble, hemem, tiered-mem-pl1, tiered-mem-pl2, panthera}. Our decision to make NVCache independent from the DRAM cache makes these techniques largely complementary. As an alternative design, a tiered memory system could be used in place of NVCache by providing a larger pool of memory into which the engine's DRAM cache could transparently expand.  Exploring this alternative was left for future work, since generic tiered memory systems known to us, e.g.,  Nimble~\cite{nimble} and HeMem~\cite{hemem} required custom kernels that were impractical do adopt in the field. CacheLib~\cite{cachelib, cache-lib-web} is a library for development of custom caches that span tiers, but as far as we understand it caches data as \texttt{items} whose raw memory can be traversed by the application, and so it would face the same need to fix pointers described in \S\ref{sec:basics} if data structures with pointers to raw memory of other items were moved between tiers. Intel Memory Mode is a tiered memory system implemented in hardware, and we compared it against NVCache in \S\ref{sec:eval}.

\section{Conclusion}\label{sec:conclusion}

Although it was well known that Optane NVRAM delivers limited write throughput, it was not known that writes disproportionately affect the throughput of reads. We discovered that in the presence of a single writer thread, the throughput of reads drops almost by a factor of 4$\times$. In contrast, with DRAM used in the same experiment the impact on read throughput was only 18\%. This discovery led us to propose a new  admission policy for Optane-resident caches. Our policy throttles the rate of writes to the cache (generated by the admission of new data, removal of invalid data and eviction), with the rate of reads, i.e., cache lookups. The metric capturing this principle, the Overhead Bypass Threshold, is generic and can be applied in any cache residing on hardware with similar properties. Our implementation outperforms an off-the-shelf cache from OpenCAS across the board, and the hardware tiered memory system (Intel Memory Mode) in all cases where the dataset size exceeds the amount of NVRAM.

\section{Availability}

The \WT source code, including NVCache, is available as open source software~\cite{wt}.

\bibliographystyle{plain}
\bibliography{biblio}

\end{document}